\documentclass[preprint,12pt]{aastex}
\usepackage{emulateapj5}

\newcommand{\kms}{\hbox{ km\thinspace s$^{-1}$}}    

\slugcomment{Scheduled for the October 2004 issue of The 
Astronomical Journal}

\shorttitle{The Multipolar Nebula Mz\,3}
\shortauthors{Guerrero et al.}

\begin{document}


\title{Mz\,3, a Multipolar Nebula in the Making}


\author{
Mart\'{\i}n A.\ Guerrero\altaffilmark{1,2,3}, 
You-Hua Chu\altaffilmark{1}, and 
Luis F.\ Miranda\altaffilmark{2}}
\affil{$^{1}$Astronomy Department, University of Illinois at 
  Urbana-Champaign, Urbana, IL 61801}
\affil{$^{2}$Instituto de Astrof\'{\i}sica de Andaluc\'{\i}a (CSIC), Spain}
\email{mar@iaa.es, chu@astro.uiuc.edu, lfm@iaa.es}


\altaffiltext{3}{
Visiting Astronomer, Cerro Tololo Inter-American Observatory, 
National Optical Astronomy Observatories, operated by the 
Association of Universities for Research in Astronomy, Inc. (AURA) 
under a cooperative agreement with the National Science Foundation.} 

\begin{abstract}

The nebula Mz\,3 has arguably the most complex bipolar morphology, 
consisting of three nested pairs of bipolar lobes and an equatorial 
ellipse.  
Its three pairs of bipolar lobes share the same axis of symmetry, but 
have very different opening angles and morphologies: the innermost 
pair of bipolar lobes shows closed lobe morphology, while the other two 
have open lobes with cylindrical and conical shapes, respectively.  
We have carried out high-dispersion spectroscopic observations of
Mz\,3, and detected distinct kinematic properties among the 
different morphological components.
The expansion characteristics of the two outer pairs of lobes suggest
that they originated in an explosive event, whereas the innermost pair 
of lobes resulted from the interaction of a fast wind with the 
surrounding material.  
The equatorial ellipse is associated with a fast equatorial outflow
which is unique among bipolar nebulae.  
The dynamical ages of the different structures in Mz\,3 suggest 
episodic bipolar ejections, and the distinct morphologies and 
kinematics among these different structures reveal fundamental 
changes in the system between these episodic ejections.  

\end{abstract}



\section{Introduction}

Mz\,3, the Ant Nebula, is perhaps one of the most stunning bipolar 
nebulae.  
The {\it Hubble Space Telescope} (\emph{HST}) color image presented by the 
Hubble Heritage Program (STScI-PRC01-05, PI: B.\ Balick, V.\ Icke, 
R.\ Sahai, and J.\ T.\ Trauger) reveals a complex system of three 
nested pairs of bipolar lobes.  
These bipolar lobes are roughly aligned along the same axis of 
symmetry, but have vastly different shapes, opening angles and 
detailed morphologies.  
In addition, a faint ellipse of emission aligned along the equator of 
these bipolar lobes is seen.
Not only is the morphology of Mz\,3 complex, but its nature is also
uncertain.   
While usually classified as a young planetary nebula (PN), Mz\,3
has also been suggested to be a circumstellar nebula of a symbiotic 
star, based on the high density of its core \citep{ZL02}, its near-IR 
colors \citep{SK01}, and the spectrum of its central star.

Previous high-dispersion spectroscopic observations of Mz\,3 have 
detected several pairs of bipolar lobes, and their kinematic properties 
led to the suggestion of episodic bipolar ejections \citep{LM83,MW85}.
More recently, \citet{Retal00} reported the discovery of fast, 500\kms,
collimated outflows. 
Detailed modeling of the structure of Mz\,3 has been hampered by the
limited detector sensitivity or sparse slit coverage of these
previous observations.
Therefore, we have carried out new long-slit, high-dispersion echelle
observations of Mz\,3, emphasizing particularly the morphological 
features that have not been observed previously.
These echelle observations, combined with the high-resolution \emph{HST} 
narrow-band images and \emph{Chandra} X-ray observation, allow us to
produce a complete spatio-kinematic model of Mz\,3, adequately
representing the three pairs of bipolar lobes and the equatorial ellipse.
While our results confirm the previous suggestion that the multipolar
structure was produced by episodic bipolar ejections \citep{LM83,MW85},
we are able to describe the kinematic properties and determine the 
formation process more precisely.
This paper reports our new observations and analysis of the physical
structure of Mz\,3.

\section{Observations}

\subsection{Archival HST Images}

Narrow-band WFPC2 images of Mz\,3 in the H$\alpha$, H$\beta$, and 
[N~{\sc ii}] $\lambda$6583 emission lines were retrieved from the 
\emph{HST} archive (Proposal IDs 6502 and 9050, PI: Balick, and 
Proposal ID 6856, PI: Trauger).  
The images we used in this work are listed in Table~1 with their 
integration times, filters, and the location of Mz\,3 on the WFPC2 
(PC or WFC).
These images were calibrated via the pipeline procedure, including 
the analog-to-digital correction, bias and dark image subtraction, 
and flat-field correction.  
We removed the cosmic rays and combined different exposures obtained 
with the same filter using standard IRAF routines.  
The H$\alpha$ and [N~{\sc ii}] images of Mz\,3, displayed in 
Figure~1, are used to analyze the nebular morphology.   
The H$\alpha$ to H$\beta$ ratio map of Mz\,3, shown in Figure~2-{\it 
left}, is used to investigate the distribution of intranebular 
extinction.

\subsection{Archival Chandra X-ray Observations}

The Advanced CCD Imaging Spectrometer (ACIS) on board the \emph{Chandra 
X-Ray Observatory} was used on  2002 October 23 to obtain a 40.8 ks 
exposure of Mz\,3 (Observation ID: 2546; PI: Kastner).  
Mz\,3 was positioned at the nominal aim point of the ACIS-S array on 
the back-illuminated S3 CCD.  
We retrieved the level 1 and level 2 processed data from the 
\emph{Chandra} Data Center and further processed the data using 
the \emph{Chandra} X-Ray Center software CIAO v3.0.2 and the 
Calibration Data Base CALDB v2.25.  
The background count rate is consistent with the quiescent 
background\footnote{Reported by M.\ Markevitch (2001), available 
at http://cxc.harvard.edu/contrib/maxim/bg/index.html.} 
for most of the observing time.
Only two background ``flares'' of short duration occurred.
After excluding these high-background periods from our analysis, 
the net exposure time was reduced to 39.6 ks.  
We used this dataset to extract an image in the 0.5-1.8 keV band at 
a resolution of $\sim$1\farcs0, and overplotted the X-ray contours 
on the \emph{HST} WFPC2 H$\alpha$ image in Figure~2-{\it right} to
illustrate the relative distribution of X-ray-emitting gas and the
ionized nebular material.

\subsection{Echelle Observations}

High-dispersion spectroscopic observations of Mz\,3 were obtained 
on 2002 June 23 and 24 using the echelle spectrograph on the CTIO 
4m telescope. 
The spectrograph was used in the long-slit mode to obtain single-order 
observations of the H$\alpha$ and [N~{\sc ii}] $\lambda\lambda$6548,6584 
lines for an unvignetted slit length of 3\arcmin.  
The 79 line~mm$^{-1}$ echelle grating and the long-focus red camera 
were used, resulting in a reciprocal dispersion of 3.4 \AA~mm$^{-1}$.  
The data were recorded with the SITe 2K No.\ 6 CCD with a pixel size of 
24 $\mu$m.  
This configuration provides a spatial scale of 0\farcs26 pixel$^{-1}$ 
and a sampling of 3.7\kms~pixel$^{-1}$ along the dispersion direction.  
The slit width was set to 0\farcs9, and the resultant instrumental 
FWHM was 8\kms.  
The angular resolution, determined by the seeing, was better than 
1\farcs2. 

The echelle observations were made with the slit oriented along 
different position angles and placed at various offsets from the 
central star, in order to sample the complex morphological features
of Mz\,3.
The slit positions and exposure times of these observations are
given in Table~2.
Although both H$\alpha$ and [N~{\sc ii}] lines are available, only the 
[N~{\sc ii}] $\lambda$6583 line is used to analyze the kinematics of 
Mz\,3 because of its smaller thermal width.
The echellograms of the [N~{\sc ii}] $\lambda$6583 line are presented 
in Figure~3, where an [N~{\sc ii}] image is also presented with the
slit positions overplotted.

\section{Results}

The \emph{HST} narrow-band images of Mz\,3 reveal three pairs of 
bipolar lobes and one elliptical feature along the equator of 
these lobes; they are marked in Figure~1 as BL1, BL2, BL3, and EE,
respectively.
These features are also detected in the echellograms and marked
correspondingly in Figure~3.  
In the following sections, we discuss detailed morphologies and 
kinematics and propose spatio-kinematical models for each of these
structures in Mz\,3.

\subsection{BL1: the Hourglass-Shaped Inner Bipolar Lobes}

The innermost pair of bipolar lobes (BL1), called Inner Bipolar 
Lobes (IBL) in \citet{Retal00}, have an hourglass morphology, with 
a narrow waist along the east-west direction (Fig.~1).  
The position-velocity diagram, i.e., the long-slit echellogram, 
also shows a tilted hourglass pattern (Fig.~3), which can be
produced by two shells expanding oppositely along the polar 
axis with the south pole tilted toward us.
The velocity difference between the walls of each lobe is 
$\gtrsim$100\kms, much larger than the $FWHM$ of the [N~{\sc ii}] 
line at the walls, 10--15\kms.  
The sharp morphology and narrow [N~{\sc ii}] line shape at the 
walls  of the BL1 lobes indicate that the material originally 
residing in the lobes has been evacuated and compressed into 
thin shells by a bipolar outflow.  
The bipolar expansion of the lobes is a direct consequence of the 
bipolarity of the outflow.
The lateral expansion of the lobes, on the other hand, may be 
driven by the thermal pressure of hot gas shock-heated by the
outflow impinging on the circumstellar material.
The hot gas in the central cavities of the BL1 lobes has been 
detected in X-rays \citep[Fig.~2-{\it right};][]{Kastner03}.

The \emph{HST} images of the BL1 lobes reveal protrusions from
their polar caps indicative of blowouts. 
The southern lobe shows a single protruding blister at its polar 
cap, while the northern lobe shows multiple blister-like structures 
extending from the polar cap and converging into a single blister 
at the end.
The [N~{\sc ii}] echellogram covering these regions, shown in 
Figure~4, reveal gas motions reflecting the blowout process.
The blister at the cap of the southern lobe shows a spindle-shaped 
[N~{\sc ii}] line that broadens up to 100 \kms\ at its leading edge,
while the multiple blister-like extensions of the northern lobe 
show multiple velocity components and the convergent blister at the 
end shows a bubble-like structure expanding rapidly both laterally
and radially.
The [N~{\sc ii}] echellograms also detect nebular knots outside the 
BL1 lobes along the axis of symmetry, as marked in Figure~4.
Exterior to the southern lobe, a bright knot at $\sim$19\arcsec\ 
from the central star is detected at roughly $-$30 km~s$^{-1}$ 
from the systemic velocity ($v_{\rm sys}$).
Exterior to the northern lobe, a counterpart of the southern knot
is detected at $\sim$19\arcsec\ from the central star with roughly 
+30 km~s$^{-1}$ offset from $v_{\rm sys}$; in addition, a fainter 
knot is detected at $\sim$25\arcsec\ from the central star with a
velocity offset of about +40 km~s$^{-1}$.

The different characteristics of the northern and southern lobes
of BL1 are probably caused by the detailed interactions between 
the bipolar outflow and the dense circumstellar material.
The circumstellar material has a high concentration in the
equatorial plane, as indicated by the higher extinction around the
waist of BL1.
The sharp band of obscuration over the northern BL1 lobe at 
$\sim2\farcs5$ north of the central star suggests that dense 
equatorial material is located in front of this lobe and therefore 
confirms the orientation of BL1 implied from its kinematics.  
The variations in the local extinction, as derived from the 
H$\alpha$/H$\beta$ ratio map shown in Figure~2-{\it left}, 
support this hypothesis:  
the H$\alpha$/H$\beta$ ratio is higher on the northern lobe than on 
the southern lobe, and therefore extinction towards the northern lobe 
is higher, indicating larger amounts of intervening material.  
In addition to the surrounding material that obscures the northern BL1 
lobe, the central star of Mz\,3 is embedded in a thick, extended shell 
detected through the mid-infrared emission of dust \citep{Quinn96}.

This H$\alpha$/H$\beta$ ratio map discloses additional clues on the 
distribution of absorbing material within Mz\,3.  
The H$\alpha$/H$\beta$ ratio, i.e., the extinction, is especially 
enhanced at the projected edge of the lobes and along the bright 
optical filaments, suggesting that the expanding lobes carry large 
amounts of dust and suffer from self absorption.    
In agreement with Smith's (2003) conclusions based on the different 
amounts of extinction derived from infrared H~{\sc i} and [Fe~{\sc ii}] 
lines, we conclude that a significant fraction of the extinction 
towards Mz\,3 is local rather than interstellar.  
The local nature of the extinction in Mz\,3 and its nonuniform 
distribution affects the morphology of the diffuse X-ray emission 
which is anticorrelated with the amount of extinction (Fig.~2), 
as typically observed in other PNe \citep{Kastner02}.

To determine the dynamical age and inclination of the polar axis
for each of the BL1 lobes, the shell morphology and position-velocity
relation need to be analyzed and modeled quantitatively.
We have adopted a simple expression to approximate the radial
expansion velocity of an hourglass as a function of the latitude 
angle, $\theta$ 
\citep{SU85}: 
\begin{equation}
v(\theta) = v_{\rm e} + (v_{\rm p} - v_{\rm e}) \times 
\sin (|\theta|)^\gamma, 
\end{equation}
where $v_{\rm e}$ and $v_{\rm p}$ are the expansion velocities 
at the equator and pole, respectively, and the exponent $\gamma$ 
sets the lobe geometry.  
We have also assumed a homologous expansion so that 
\begin{equation}
r(\theta) = \Delta t \times v(\theta)
\end{equation}
where $\Delta t$ is the time since the lobe was formed.

Using the model outlined above, we have determined $v_{\rm e}$ and 
$v_{\rm p}$, the age, the exponent $\gamma$, and the inclination 
with respect to the sky and PA of the symmetry (polar) axis of each 
of these bipolar lobes.  
The best fits for the southern and northern BL1 lobes are shown 
in Figure~5 and the parameters of these fits are listed in Table~3.  
As expected from the different morphological and kinematical properties 
of the southern and northern lobes, the best-fit parameters to each 
lobe are not exactly the same, though both fits have similar inclination 
of the symmetry axis with respect to the plane of the sky, 
15\arcdeg--20\arcdeg, and kinematical age\footnote{
The distance to Mz\,3 is highly uncertain.  
Hereafter we have chosen to show explicitly the dependence of the 
kinematical age on distance. }, 
(600$\pm$50)${\times}(\frac{D}{\rm kpc})$ yr, where $D$ is the distance 
in kpc to Mz\,3.  

If the bright knots at the tip of the bipolar lobes share their 
inclination angle, then the true de-projected velocity of these 
knots is in the range between 90 km~s$^{-1}$ and 150 km~s$^{-1}$.  
For comparison, we have also included in Tab.~3 the parameters of 
the best fit to the northern lobe considering the extension and 
kinematics of the converging blister at its polar cap.  
The shorter kinematical age of the northern lobe when its blister 
is considered may be suggestive of acceleration of the gas motions 
caused by a blowout process.

\subsection{BL2: the Cylindrical Bipolar Lobes}

The bipolar lobes of BL2, the Outer Bipolar Lobes 1 (OBL1) in 
\citet{Retal00}, have an almost rectangular morphology, with the PA's 
of the western and eastern edges having a difference as small as 
$\sim$5\arcdeg\ (Fig.~1).  
The lobes have a width of $\sim$24\arcsec\ and a length up to 
$\sim$85\arcsec\ for the northern BL2 lobe, i.e., the aspect 
ratio is 7:1.  
Their edges are rather straight, bending inwards only at the 
location where these lobes contact the inner BL1 lobes.  
The detailed morphology of BL2 shows a complex system of long filaments 
extending radially outwards.  
These filaments originate from a collection of knots at the base of 
the BL2 lobes that form a cavity-like structure just outside the BL1 
lobes (Fig.~4).

The [N~{\sc ii}] echellogram of the BL2 lobes along PA 8\arcdeg\ 
(i.e., roughly the BL2 symmetry axis) shows two velocity components 
with a velocity gradient of $\sim$1.1\kms~arcsec$^{-1}$ (Fig.~2).  
The difference in velocity between these two components, $\sim$110\kms, 
does not change significantly with the position along the symmetry 
axis of these lobes.  
Along the orthogonal direction, the echellograms at PA 98\arcdeg\ 
and offset 14\arcsec, 19\arcsec, and 26\arcsec\ show hollow 
position-velocity ellipses in BL2 (Fig.~2).  
Material in these lobes is thus mostly confined in the thin walls of 
hollow cylinders.  
This material cannot just flow along the walls of the cylinder, as 
a cross section of such a cylinder would have exactly the same 
observed velocity.  
Instead, the apparently constant velocity-split implies that 
the section of the cylinder expands with a constant, $\sim$55\kms, 
expansion velocity.  
Meanwhile, the velocity along the walls increases with the distance 
from the central star and must be faster than the transversal 
velocity; 
otherwise the lobes will not show the high aspect ratio, $\sim$7:1, 
that characterizes them.

The Hubble law-like expansion of the BL2 lobes suggests that 
these lobes were made in a single, explosive event.  
As the difference in velocity between the blue- and red-shifted 
components at a given location of BL2 is the same, $\sim$110\kms, 
the kinematical age of BL2 can easily be derived assuming that 
the 24\arcsec\ width of BL2 are simply due to expansion along 
this direction.  
The kinematics are well reproduced using a cylinder\footnote{
Actually, these are not cylinders, as they are opening gradually 
with distance from the central star, but the angle of divergence, 
$\sim$5\arcdeg, is too small to affect significantly the model fits.}  
tilted with respect to the plane of the sky with fixed, 55\kms, 
expansion velocity across its section and linearly increasing 
velocity along the walls (Figure~6).  
The best fit model has an inclination of 20\arcdeg$\pm$5\arcdeg\ 
against the sky plane, in agreement with the previous value reported 
by \citet{MW85}, and a kinematical age of 
(1,000$\pm$100)${\times}(\frac{D}{\rm kpc})$ yr.  
At the maximum distance of 85\arcsec\ from the central star of Mz\,3, 
the de-projected expansion velocity is $\sim$320\kms.

\subsection{BL3: the Conical Lobes}

We have named BL3 the pair of bipolar lobes with conical shape, 
called the Outer Bipolar Lobes 2 (OBL2) by \citet{Retal00}.  
These lobes have an opening angle of $\sim$50\arcdeg\ and their 
limbs point directly to the central star of Mz\,3. 
In the images in Fig.~1, the conical lobes BL3 are composed of 
multiple knots with long, radial tails stretching out up to 
60\arcsec\ from the central star of Mz\,3.  
The distribution of these knots and filaments is looser than this 
of the filaments in  BL2.  
Indeed, the knots and filaments in BL3 look disconnected, more like 
individual streams of material than as part of a contiguous structure.

Further information on the kinematics and structure of BL3 can be 
derived from the echellograms at PA 43\arcdeg, 52\arcdeg, and 
$-$28\arcdeg\ through the central star, and at PA 98\arcdeg\ and 
offset 14\arcsec, 19\arcsec, and 26\arcsec\ to the south of 
the central star of Mz\,3 (Figs.~3 and 7).  
In the echellograms at PA's 43\arcdeg, 52\arcdeg, and $-$28\arcdeg, 
the knots and filaments composing BL3 appear as tilted straight 
features with different slopes on the position-velocity space.  
The structure and kinematics of BL3 in these echellograms is somehow 
confused by that of BL2, but the echellograms at PA 98\arcdeg\ 
resolve unambiguously BL3 from BL2. 
In these echellograms, the velocities of the knots and filaments 
of BL3 are mostly distributed, but not completely confined, along 
ellipses.  
The radial velocities of these ellipses as well as the velocity 
differences between their red- and blue-shifted sides increase 
radially from the central star of Mz\,3.  
The distribution in the position-velocity space of these knots 
and filaments suggests that, unlike BL2, material in BL3 is not 
completely confined to the walls of the conical lobes.  
This is illustrated by the feature seen in the echellograms at PA 
98\arcdeg\ and offset 14\arcsec\ South and 19\arcsec\ South at 
relative position $\sim-$15\arcsec\ and 
$(v-v_{\rm sys})\sim-$100\kms\ (Figs.~2 and 7).  
This feature looks like a small velocity ellipse whose spatial size 
and difference in velocity increase from the echellogram at offset 
14\arcsec\ South to that at 19\arcsec\ South, suggesting that this 
filament is opening into a conical structure.

The kinematics of the knots and filaments of BL3 derived from these 
echellograms show that their expansion velocity follows a Hubble law.  
We have modeled the kinematics of the BL3 lobes assuming that 
they have a conical shape with opening angle $\sim$50\arcdeg, 
and that the expansion velocity is directed along the walls 
of the cone and increases linearly outwards from the central 
star of Mz\,3. 
Following this model, we have fit the observed kinematics (Fig.~7) and 
derived an inclination angle of the symmetry axis with the plane of the 
sky of 12\arcdeg$\pm$3\arcdeg.  
The de-projected expansion velocity at 60\arcsec\ from the central 
star would be 180$\pm$30\kms\ and the kinematical age of BL3 is 
(1,800$\pm$200)${\times}(\frac{D}{\rm kpc})$ yr.  
The inclination angle and kinematical age derived from this fit have 
greater uncertainty than these fitting BL1 and BL2, because the 
discrete nature of BL3 makes difficult to judge the goodness of the 
fit and to determine the best-fit parameters.

\subsection{EE: the Equatorial Ellipse} 

The \emph{HST} images of Mz\,3 displays an additional feature 
unnoticed in previous images, a closed ellipse with size 
82\arcsec$\times$32\arcsec\ oriented along PA$\sim$85\arcdeg, 
i.e., almost along the nebular equator (Fig.~1).  
This structure, referred to as the Equatorial Ellipse (EE), is 
delineated by filamentary arcs especially prominent at the 
northeast and southwest of Mz\,3.

EE is revealed as dramatic high-velocity arcs in the echelle 
observations along the slits oriented at PA 98\arcdeg\ and offsets 
3\arcsec\ North, and 4\arcsec, 14\arcsec, and 19\arcsec\ South of 
the central star, as well as in the slits at PAs 8\arcdeg, 52\arcdeg, 
43\arcdeg, and $-$28\arcdeg\ (Fig.~3).  
The measured expansion velocity is close to 200\kms\ with respect 
to the systemic velocity.   
It is interesting to note that the arcs in the echellograms of 
the slits passing through the central star are disrupted by radial 
filaments of BL3.  
It is also interesting to note that the arcs detected in the 
echellograms at PA 98\arcdeg\ and offsets 3\arcsec\ North and 
4\arcsec\ South show marked point-symmetry.

It is clear from these results that Mz\,3 shows an equatorial outflow 
moving at high velocity.  
Its three-dimensional geometry, however, is difficult to envision 
because the fragmented information revealed by the observations 
and the likely interaction of EE with BL3.  
In the following, we will consider four different geometrical models
for this outflow: 
(a) an extended equatorial disk, 
(b) a ring collimating a bipolar ejection, 
(c) a pair of wide-opened bipolar lobes, and 
(d) an oblate ellipsoid-like shell.

Although equatorial disks have been proposed to play an important 
role in the collimation of bipolar PNe, there is no detection of 
high velocity equatorial disks in PNe.  
An example of equatorial disk can be found in the bipolar nebula around 
$\eta$ Carinae which shows an equatorial structure that seems to be an 
extended equatorial disk (Smith 2002).  
If EE in Mz\,3 were a circular equatorial disk, then the expansion law 
with radius on the disk can be inferred from any echellogram of a slit 
passing through its center, simply by applying a scaling factor that depends 
on the inclination of the disk with respect to the line of sight, 
because all velocities along such a line share the same inclination 
angle with the line of sight.  
For a circular disk, its inclination angle can be derived from 
the observed minor-to-major axes ratio of the projected ellipse.  
The size of EE of 82\arcsec$\times$32\arcsec\ corresponds to 
an inclination against the plane of the sky of the rotation 
axis of the circular disk of $\sim$23\arcdeg.  
Using this value for the inclination of the disk and the information on 
the expansion velocity law on the slit at PA 52\arcdeg\ passing through 
the center of Mz\,3, we have determined the expansion law with radius 
on the disk which is plotted in Figure~8.  
The velocity in the disk decreases smoothly with radius up to a given 
radius, when the velocity decreases sharply.  
Using this velocity law, we have produced the position-velocity plots 
expected for significative slit positions (Fig.~9).   
The model deviates significantly from the position-velocity arcs 
observed along PA 98\arcdeg\ with different offsets from the central 
star (Fig.~9). 
We conclude that EE cannot be interpreted as an expanding disk.

A detailed study of the spatio-kinematical properties of an expanding 
ring collimating a pair of bipolar lobes is presented by \citet{SU85}  
for the bipolar nebula around the symbiotic Mira variable R Aqr.  
In an expanding ring, the ring itself projects an ellipse onto the sky, 
long-slit echellograms along the ellipse major axis show two arcs in 
the position-velocity space, one shifted to the blue and the other to 
the red, and long-slit echellograms along the ellipse minor axis reveal 
a characteristic hourglass shaped line.  
The morphology and kinematics of Mz\,3 observed in the echellograms of 
the slits along PA 98\arcdeg\ are compatible with this model 
expectations; however, the slit at PA 98\arcdeg\ and offset 
26\arcsec\ South of the central stars does not detect emission 
outside the observed ellipse, nor the slits at PAs 52\arcdeg, 
43\arcdeg, 8\arcdeg, and $-$28\arcdeg\ show the expected hourglass 
shape.  
We conclude that an expanding ring that collimates bipolar lobes is not 
appropriate for the three-dimensional geometry of EE.

We have also considered the possibility that EE is composed by a 
pair of wide-opened, champagne-glass-shaped bipolar lobes tilted 
with the line of sight so that the flow vector points almost
directly to us at the location of the equatorial waist.  
This model would explain the observed kinematics: 
at locations near the central star, the observed velocity is large
because the line of sight is close to the direction of the flow 
vector, while at increasing distances from the central star, the 
lobes bend and close so that the direction of the flow vector 
diverges from the line of sight and the observed velocity decreases.  
The projection of these lobes onto the plane of the sky, however, 
would not produce an elliptical shape, but two interwined arcs 
pointing at opposite directions, as observed, e.g., in the central 
regions of MyCn\,18 \citep{Sahai99}.  
We thus disregard this model as the three-dimensional geometry of EE.

Finally we consider an oblate shell which expands much faster along 
the equator than along the poles and whose symmetry axis is close 
to the plane of the sky.  
Assuming a homologous expansion for this shell, we have produced 
synthetic position-velocity plots that can be compared with the 
observed ones to determine the best fit parameters (Figure~10).  
The best-fit shell model is a flat ellipsoid-like whose symmetry axis 
is tilted against the line of sight by 70\arcdeg$\pm$5\arcdeg, and the 
ellipsoid-like has an equatorial expansion velocity $\simeq$200\kms, a 
polar velocity $\simeq70\pm$20\kms, and a kinematical age 
(1,000$\pm$50)${\times}(\frac{D}{\rm kpc})$ yr.  
This model explains satisfactorily the high velocity arcs observed in 
the slits at PA 98\arcdeg.  
It also accounts for the disruption of EE by BL3, which has bored a hole 
near the polar regions of EE.  
Finally, this model also explains the point-symmetric distribution of 
arcs observed in the slits at PA 98\arcdeg\ and offsets 3\arcsec\ North 
and 4\arcsec\ South;  
at these locations, the detectability of the shell is optimized because 
the shell is seen tangentially and the optical path is thus larger than 
at other locations.

\section{The Multipolar Structure of Mz\,3}

Previous spatio-kinematical studies of Mz\,3 have revealed an 
increasing level of complexity in this nebula.    
\citet{LM83} studied the inner bipolar lobes (BL1) and concluded that 
these lobes are hourglass in shape with the symmetry axis close to 
the line of sight.  
In a later paper, \citet{MW85} determined with greater accuracy a 
spatio-kinematical model of the inner bipolar lobes.  
Moreover, they extended the spatio-kinematical study of Mz\,3 to 
the outer regions, reporting the presence of different sets of bipolar 
lobes.  
The low spatial resolution of the narrow-band images available by then, 
however, hampered Meaburn \& Walsh's study:  
the bipolar lobes BL2 and BL3 were not distinguished from each 
other; the equatorial ellipse was interpreted as an additional 
bipolar lobe; and the detection of a high-velocity component in 
the Na\,{\sc i} line, correctly interpreted as related to a 
high-velocity outflow from Mz\,3, 
was not associated to the equatorial ellipse EE.  
More recently, \citet{Retal00} obtained high-dispersion spectroscopic 
observations along the major axis of Mz\,3 that allowed them to 
describe the kinematics of the bipolar lobes BL2 and to find 
high-velocity, $\sim$200\kms, components at the location of the
blowout at the tips of the inner bipolar lobes BL1.  
Because of the limited spatial coverage of their study, the
association between these high-velocity kinematical components 
and the equatorial ellipse EE was not as clearly seen as 
evidenced in our echelle observations obtained at different 
slit positions (Fig.~3). 
Finally, in a simultaneous study of Mz\,3, \citet{SG04} have derived 
spatio-kinematical models and kinematical ages for the three pairs 
of bipolar lobes that are in complete agreement with these derived 
here.

The present study reconciles many of the previously reported
kinematical features of Mz\,3 into a more comprehensive view of 
its physical structure.  
Mz\,3 consists of four distinct structures, an oblate ellipsoid-like 
shell and three pair of bipolar lobes with almost coincident 
symmetry axes.  
The properties of these structures are especially singular among 
similar structures observed in bipolar PNe.  
Unlike the slowly expanding rings or tori observed in some bipolar 
PNe, the oblate ellipsoidal-like shell expands at high velocity 
along the equator of the bipolar lobes.  
Similarly, very few multipolar PNe have pairs of lobes exhibiting 
the notable differences in opening angle, morphologies and detailed 
small-scale structures as the three pairs of bipolar lobes of Mz\,3; 
BL1 has hourglass-shaped expanding bubbles filled with 
X-ray-emitting hot gas (Kastner et al.\ 2003), while BL2 
and BL3 are composed of knots and filaments following a 
Hubble flow with cylindrical and conical shapes, 
respectively.

Multipolarity has become a common feature among bipolar nebulae.  
A growing number of bipolar nebulae have been noted to have multiple 
systems of bipolar lobes either sharing the same symmetry axis or 
having different symmetry axes, e.g., M\,2-9, M\,2-46, NGC\,2440, 
and Hen\,2-104 \citep{HL94,MGS96,LMBH98,Solf00,Corradi01}.  
Among these multipolar nebulae, the case of Mz\,3 is of especial 
interest because the kinematical ages of the different systems of 
bipolar lobes in Mz\,3 are small and of the order of the difference 
in kinematical ages among them.  
The inner bipolar lobes BL1 have a kinematical age 
500--600${\times}(\frac{D}{\rm kpc})$ yr, 
the cylindrical lobes BL2 and the equatorial ellipsoid EE are 
$\sim$1,000${\times}(\frac{D}{\rm kpc})$ yr old, and 
the outer bipolar lobes BL3 have a somewhat more uncertain kinematical 
ages of $\sim$1,800${\times}(\frac{D}{\rm kpc})$ yr.  
Mz\,3 is thus a multipolar nebula in the making, where BL1 corresponds 
to the most recent ejection from Mz\,3 central star, EE and BL2 are 
older and probably coeval, and BL3 is finally the oldest structure, 
although its kinematical age is the most uncertain and we cannot rule 
out a formation closer in time to that of BL2.

The different kinematical properties of the three pairs of bipolar 
lobes suggest distinct formation scenarios.  
The ballistic motion of the two outermost bipolar lobes of Mz\,3, BL2 
and BL3, indicates that the gas within these lobes expands freely under 
its own inertia.  
Most likely, these lobes are the result of two episodes 
of explosive mass ejection or outbursts that occurred 
$\sim$1,800${\times}(\frac{D}{\rm kpc})$ yr and 
$\sim$1,000${\times}(\frac{D}{\rm kpc})$ yr ago.  
The last episode of mass ejection responsible of BL2 also resulted in 
high velocity ejecta along the equatorial plane that formed EE, the 
equatorial ellipse.  
On the other hand, the morphology and hot gas content of the innermost 
pair of lobes, BL1, indicate that they resulted from the interaction of 
highly pressurized hot gas with the surrounding material.  
This hot gas may be produced by the onset of a fast stellar wind.  
An alternative origin has been proposed by \citet{Kastner03} who 
attribute the X-ray emission to the action of an X-ray jet along the 
symmetry axis of Mz\,3.  
Our observations indeed reveal bipolar collimated outflows along 
the symmetry axis of Mz\,3 (the knots further away the leading 
edges of the BL1 lobes as seen in Fig.~4), but not with the high 
velocities required to produce the observed X-ray emission.  
Note, however, that the outflow detected in our observations may 
trace high density material accelerated by a much higher velocity 
jet that, being responsible of the X-ray emission, would elude 
optical detection because its low density.

The oblate shell forming the equatorial ellipse EE of Mz\,3 is a 
very singular structural component.  
Many bipolar PNe show equatorial disks or tori, but all of them have 
modest expansion velocities, $\sim$30\kms.
Bipolar nebulae around symbiotic stars also show equatorial disks 
or tori, but expansion velocities are modest, too.  
The only exception among symbiotic stars is the remarkable 
elliptical shell or ring around Hen\,2-147 with an expansion 
velocity $\sim$100\kms\ \citep{Corradi99}.  
Thus, the $\ga$200\kms\ equatorial outflow in Mz\,3 is the most
extraordinary among bipolar PNe and nebulae around symbiotic 
stars.

The equatorial outflow of Mz3 rivals that of the massive star
$\eta$ Car.
The equatorial outflow around $\eta$ Car shares many similarities with 
this found in Mz\,3:  
the nebula around $\eta$ Car has several systems of bipolar lobes 
\citep{Ishibashi03} and the formation of the equatorial outflow has 
been timed during or around the moment when the main bipolar lobes 
in $\eta$ Car, the Homunculus Nebula, were formed.  
Despite these similarities, the equatorial outflows in both nebulae 
are notably different. 
The equatorial outflow in $\eta$ Car has been described as an 
extended equatorial disk expanding with velocity proportional 
to the angular distance to center \citep{Davidson01,Smith02}, 
while the physical structure of the equatorial outflow in 
Mz\,3 is best described by an oblate shell.  
Furthermore, their detailed morphologies are different and very 
likely indicate different origins:  
in $\eta$ Car, the equatorial outflow seems to be composed of multiple 
jet-like features located along the equatorial plane, while in 
Mz\,3 the equatorial outflow shows the limb-brightened morphology 
characteristic of a thin shell.

The formation of multipolar nebulae can be explained as the result 
of recurrent outbursts as those observed in massive stars in binary 
systems during the Luminous Blue Variable (LBV) phase, e.g. $\eta$ 
Car.  
In low mass stars, recurrent outbursts can be related to nova-like 
eruptions on the accreting hot component of a symbiotic star or 
to structural instabilities in the late evolution of the central 
star of a PN (e.g., thermal pulses).  
In symbiotic novae, the timescales of successive outbursts are 
determined by the mass of the accreting white dwarf, the mass 
loss rate of the red giant, and the accretion efficiency of the 
wind capture which is related to the binary interaction 
\citep[e.g.][]{PK95}.    
Recurrence periods of a few hundred years are typical of symbiotic 
novae \citep{PK95,Corradi99}.  
The formation of multipolar PNe is far more difficult to explain, as 
it requires the alternation between a dense, slow wind and a fast, 
tenuous wind.  
The evolution of the central star of the PN in a binary system provides 
a natural scenario for recurrent outbursts during the evolution through 
a common envelope phase or as the result of accretion and nova-like 
outbursts on the white dwarf component of a symbiotic star.  
This raises the similarities between Mz\,3 and other symbiotic stars 
like R~Aqr and Hen\,104, or other suspected symbiotic stars yet 
classified as PNe, e.g., M\,2-9, and casts doubts on the true nature 
of Mz\,3 as a PN.

Even if we accept that Mz\,3 has formed as the result of recurrent 
nova-like outbursts in a symbiotic star, the physical structure of 
this bipolar nebula is rather unique.  
The successive collimated ejections in Mz\,3 are rather regular 
in time, but they have very different morphological and kinematical 
properties, which suggest very distinct conditions and formation 
mechanisms. 
In Mz\,3, we are thus witnessing the formation of a multipolar nebula 
which evolves dramatically between periodic outburst episodes.

\acknowledgments

M.A.G. and L.F.M. acknowledge support from the grant AYA~2002-00376 of 
the Spanish MCyT (cofunded by FEDER funds).  
We thanks Miguel Santander Garc\'{\i}a for providing us with the 
results on their spatio-kinematical modeling of Mz\,3 before 
publication.  
We also thank the referee, Dr.\ Matt Redman, for his valuable comments.

\clearpage

\begin{figure}
\caption{
\emph{HST} WFPC2 images of Mz\,3 in the H$\alpha$ ({\it top}) and 
[N~{\sc ii}] $\lambda$6583 ({\it bottom}) emission lines.  
The different morphological components of this nebula are marked as 
described in the text:  
BL1 are the inner bipolar lobes, BL2 the cylindrical lobes, BL3 the 
conical lobes, and EE is the  equatorial ellipse.  
Both the H$\alpha$ and [N~{\sc ii}]  images are displayed at two 
intensity contrasts and spatial scales to highlight these different 
morphological components.  
}
\end{figure}

\begin{figure}
\caption{
{\it Left:} H$\alpha$ to H$\beta$ intensity ratio map of Mz\,3.  
Bright regions correspond to higher H$\alpha$/H$\beta$ ratio and thus 
to higher extinction.  
{\it Right:} \emph{HST} WFPC2 H$\alpha$ image of Mz\,3 overlaid with 
X-ray contours in the 0.5--1.8 keV energy range extracted from a 
\emph{Chandra} ACIS-S observation.   
}
\end{figure}

\begin{figure}
\epsscale{0.85}
\caption{
[N~{\sc ii}] $\lambda$6583 image ({\it top-left}) and echellograms of 
Mz\,3 along 9 different slit positions.  
The slits positions of the echelle observations are plotted over the 
[N~{\sc ii}] image.  
The different morphological components of this nebula are marked on 
the echellograms.  
Note that the spatial scale of the image and echellograms are not 
coincident.  
Note also that velocities have been referred to the systemic velocity 
of Mz\,3.  
}
\end{figure}

\begin{figure}
\epsscale{0.60}
\caption{
[N~{\sc ii}] $\lambda$6583 image ({\it left}) and echellogram along PA 
8\arcdeg ({\it right}) of Mz\,3.  
Both the image and echellogram have the same orientations and spatial 
scales to make easy a fair comparison.  
The arrows indicate the locations in the image and the echellogram of 
the knots at the tips of the innermost bipolar lobes described in the 
text.  
Contrast in the image has been chosen to highlight these features.  
}
\end{figure}

\begin{figure}
\caption{
[N~{\sc ii}] $\lambda$6583 image ({\it left}) and echellogram along PA 
8\arcdeg ({\it right}) of Mz\,3 overlaid by the best model fit for the 
hourglass bipolar lobes described in $\S3.1$.  
Image and echellogram are shown at the same spatial scale.  
}
\end{figure}
\begin{figure}
\epsscale{0.70}
\caption{
[N~{\sc ii}] $\lambda$6583 echellograms of Mz\,3 along selected slit 
positions marked on the figure.  
The echellograms are overlaid by the position-velocity plots derived 
from the best model fit for the cylindrical bipolar lobes BL2 
described in $\S3.2$.  
}
\end{figure}

\begin{figure}
\epsscale{0.60}
\caption{
[N~{\sc ii}] $\lambda$6583 echellograms of Mz\,3 along selected slit 
positions marked on the figure.  
The echellograms are overlaid by the position-velocity plots derived 
from the best model fit for the conical bipolar lobes BL3
described in $\S3.3$.  
}
\end{figure}

\begin{figure}
\epsscale{1.00}
\caption{
Radial dependence of the expansion velocity of an expanding equatorial 
disk as inferred from the information of the EE component in the 
echellogram along PA 52\arcdeg.  
The circular disk has an inclination 23\arcdeg\ consistent with the 
minor-to-major axes ratio of EE.  
}
\end{figure}

\begin{figure}
\epsscale{0.60}
\caption{
[N~{\sc ii}] $\lambda$6583 echellograms of Mz\,3 along selected slit 
positions marked on the figure.  
The echellograms are overlaid by the position-velocity plots derived 
from the best model fit for an expanding equatorial disk with the 
expansion law given in Figure~8.  
This best model fit reproduces the kinematics of EE seen in the 
echellograms for slit positions passing through the nebular center, 
but the fit is very poor for the slit positions offset from the 
center and at directions orthogonal to the projected symmetry axis 
of the nebula.  
}
\end{figure}

\begin{figure}
\epsscale{0.9}
\caption{
[N~{\sc ii}] $\lambda$6583 echellograms of Mz\,3 along selected slit 
positions marked on the figure.  
The echellograms are overlaid by the position-velocity plots derived 
from the best model fit for an oblate shell as described in $\S3.4$.  
This model produces an acceptable fit for all slit positions.  
}
\end{figure}

\clearpage

\begin{deluxetable}{lrcl}
\tablenum{1}
\tablewidth{0pt}
\tablecaption{Archival \emph{HST} WFPC2 Observations of Mz\,3}
\tablewidth{0pt}
\tablehead{
\multicolumn{1}{c}{Emission Line} & \multicolumn{1}{c}{Exposure Time} & 
\multicolumn{1}{c}{Location} & \multicolumn{1}{c}{Program ID} \\
\multicolumn{1}{c}{} & \multicolumn{1}{c}{(sec)} & 
\multicolumn{1}{c}{PC1/WF3} & \multicolumn{1}{c}{}
}
\startdata
~~~~~~H$\alpha$     &  350~~~~ & PC1 & ~~~9050 \\
~~~~~~H$\alpha$     &  900~~~~ & WF3 & ~~~6856 \\
~~~~~~H$\beta$      & 1300~~~~ & WF3 & ~~~6502,6856 \\
~~~~~~[N~{\sc ii}]  & 1300~~~~ & PC1 & ~~~9050 \\ 
~~~~~~[N~{\sc ii}]  &  900~~~~ & WF3 & ~~~6856 \\
\enddata
\end{deluxetable}

\begin{deluxetable}{rrr}
\tablenum{2}
\tablewidth{0pt}
\tablecaption{Echelle Observations}
\tablewidth{0pt}
\tablehead{
\multicolumn{1}{c}{Offset} & \multicolumn{1}{c}{Position Angle} & 
\multicolumn{1}{c}{Exposure Time} \\
\multicolumn{1}{c}{(\arcsec)} & \multicolumn{1}{c}{(\arcdeg)} & 
\multicolumn{1}{c}{(sec)}  
}
\startdata
   0~~~~ &    8~~~~~~~~~ &  600~~~~~~~~ \\
   2 W   &   43~~~~~~~~~ &  900~~~~~~~~ \\
   0~~~~ &   52~~~~~~~~~ &  900~~~~~~~~ \\
   3 N~  &   98~~~~~~~~~ & 1800~~~~~~~~ \\
   4 S~  &   98~~~~~~~~~ & 1800~~~~~~~~ \\
   8 S~  &   98~~~~~~~~~ & 1800~~~~~~~~ \\
  14 S~  &   98~~~~~~~~~ & 1800~~~~~~~~ \\
  19 S~  &   98~~~~~~~~~ & 1800~~~~~~~~ \\
  26 S~  &   98~~~~~~~~~ & 1800~~~~~~~~ \\
   0~~~~ &$-$28~~~~~~~~~ &  900~~~~~~~~ \\
\enddata
\end{deluxetable}

\begin{deluxetable}{lrrrrr}
\tablenum{3}
\tablewidth{0pt}
\tablecaption{Fits of the Physical Parameters of BL1} 
\tablewidth{0pt}
\tablehead{
\multicolumn{1}{c}{Bipolar Lobe} & 
\multicolumn{1}{c}{$v_{\rm p}$}  & 
\multicolumn{1}{c}{$v_{\rm e}$}  & 
\multicolumn{1}{c}{$i$}          & 
\multicolumn{1}{c}{PA}           & 
\multicolumn{1}{c}{Kin. Age}     \\
\multicolumn{1}{c}{}             & 
\multicolumn{1}{c}{(km~s$^{-1}$)} & 
\multicolumn{1}{c}{(km~s$^{-1}$)} & 
\multicolumn{1}{c}{(\arcdeg)} & 
\multicolumn{1}{c}{(\arcdeg)} & 
\multicolumn{1}{c}{(yr)}  
}
\startdata
Southern Lobe             & 100~~~~ &  15~~~~ & 15 &  8 & 600~~~~ \\  
Northern Lobe             &  80~~~~ &  15~~~~ & 20 & 12 & 600~~~~ \\  
Northern Lobe and Blister & 140~~~~ &  15~~~~ & 20 & 10 & 520~~~~ \\  
\enddata
\end{deluxetable}

\end{document}